\documentstyle[aps,prb,eqsecnum,twocolumn,psfig]{revtex}
\begin{document}
\draft 

\twocolumn[\hsize\textwidth\columnwidth\hsize\csname
@twocolumnfalse\endcsname

\title{Density-functional theory of freezing of quantum liquids at
zero temperature using exact liquid-state linear
response }
	   
\author{C. N. Likos,$^{1,}$\cite{email1}
Saverio Moroni,$^{2,}$\cite{email2} and Gaetano Senatore$^{1,}$\cite{email3}}
\address{ $^{1}$Dipartimento di Fisica Teorica dell'Universit\`a di
Trieste and Istituto Nazionale di Fisica della Materia, \\ Strada
Costiera 11, I-34014 Grignano, Trieste, Italy\\ $^{2}$International
Centre for Theoretical Physics, Strada Costiera 11, I-34014 Trieste,
Italy } \date{Submitted to Physical Review B, August 27, 1996} \maketitle

\begin{abstract}
We apply density functional theory to study the freezing of superfluid
{$^{4}\rm{He}$}, charged bosons and charged fermions at zero
temperature. We employ accurate Quantum Monte Carlo data for the
linear response function in the uniform phase of these systems, a
quantity that has different behavior for large values of the
wavevector than previously assumed. We find that, as a result of this
{\it{exact}} behavior, different approximations in the density
functional theory of freezing that involve linear response, all fail
to correctly describe the crystallization in {\it{three dimensions}},
while yielding satisfactory predictions in {\it{two dimensions}}.
This demonstrates the shortcomings of the currently popular density
functional approximate theories to describe $3d$-freezing in the
quantum regime.  We also investigate the consequences of the exact
asymptotic behavior of response functions on the form of effective
interactions and polarization potentials in the electron gas, at small
distances.
\end{abstract}
\vskip2pc]

\renewcommand{\thepage}{\hskip 8.9cm \arabic{page} \hfill Typeset
using REV\TeX }

\narrowtext

\section{Introduction}

The modern density functional theory (DFT), which is employed in the
theoretical investigations of freezing of both quantal and classical
systems, is based on an exact correspondence between equilibrium
one-particle densities and external potentials.\cite{kohn,mermin} In
particular, if we denote by $n({\bf{r}})$ the one-particle density of
the system (i.e. the statistical average of the one-particle density
operator) the system can be characterized by an appropriate
thermodynamic potential which attains its minimum value for the
correct (equilibrium) profile $n_0({\bf{r}})$. For the study of
crystallization, the relevant thermodynamic potentials are the grand
potential $\Omega$ and the intrinsic Helmholtz free energy $F$, the
latter being a {\it{unique functional}} of the one-particle
density.\cite{kohn,mermin} If $\mu$ is the chemical potential of the
system at some temperature $T$ and $v_{ext}({\bf{r}})$ is an arbitrary
external potential, then the quantity:
\begin{eqnarray}
 \tilde\Omega[n,u] = F[n]-\int d{\bf{r}} n({\bf{r}}) u({\bf{r}}), 
 \label{var} 
\end{eqnarray}
where $u({\bf{r}}) = \mu - v_{ext}({\bf{r}})$, is a {\it{minimum}} for
given $u({\bf{r}})$ at the equilibrium density $n_0({\bf{r}})$. The
quantity $\Omega[u] = \tilde\Omega[n_0,u]$ is then the grand potential
of the system. Clearly, the equilibrium condition reads as
\begin{eqnarray}
 {{\delta F[n]}\over{\delta n({\bf{r}})}}\Bigg|_{n_0({\bf{r}})} =
 u({\bf{r}}). \label{equil} 
\end{eqnarray}
For vanishing external potential and fixed particle number $N$, the
intrinsic free energy $F[n]$ is a minimum at the equilibrium density,
with respect to variations of the {\it{shape}} of the density
profile. It is customary to separate $F[n]$ into a contribution from
the noninteracting system under a suitable external potential that
makes $n({\bf{r}})$ the equilibrium density, $F_{id}[n]$, and an
excess part $F_{ex}[n]$, i.e.  $F[n] = F_{id}[n] + F_{ex}[n]$.  The
determination of $F_{id}$ is not complicated, for both classical and
quantum systems: in the former case, $F_{id}$ is known explicitly as a
functional of the density.\cite{evans} In the latter case, the
statistics appears explicitly in the construction of $F_{id}$, and for
given external potential one can construct both the equilibrium
density and $F_{id}$ in a straightforward manner. Therefore, the art
of density functional theory amounts to the invention of approximate
functionals for the excess part. In the classical regime, there has
been extensive work in this direction during the last fifteen years.
\cite{reviews} Relatively less has been done in the quantum regime,
with which we are concerned in this work.

The development of quantum DFT of freezing has followed two
alternative routes: In one case,\cite{haymet} suitable for finite
temperatures, a mapping of the quantum particles into classical
polymer rings is invoked; in the other, which is better suited for
zero temperature, the Hohenberg-Kohn-Sham formalism\cite{kohn,sham} is
used, and the problem is reduced to a self-consistent band structure
calculation.\cite{senatore,denton,saverio} Here we follow the second
approach, since we are interested in $T=0$ freezing. In this case,
$F[n]$ is simply the intrinsic ground-state energy $E[n]$. The ideal
part $F_{id}[n]$ reduces to the kinetic energy of noninteracting
particles $T_0[n]$, and the remainder $F_{ex}[n]$ is the excess energy
$E_{ex}[n]$. A brief summary of this formalism will be presented
below.

Within certain classes of approximate functionals, an essential
ingredient for the practical implementation of this approach is the
linear response function $\chi(r;n)$ of the fluid, or its Fourier
transform $\tilde\chi(q;n)$ where $q$ is the wavevector magnitude and
$n$ is the average density. In particular, what is important is the
`quantum' direct correlation function (dcf), i.e., {\it{the
difference}} between the inverse linear response functions of the
interacting and noninteracting systems, $\tilde
K(q,n)=\tilde\chi^{-1}(q;n)-\tilde\chi_0^{-1}(q;n)$.  In previous
applications\cite{senatore,denton,saverio,moroni} it was assumed that
this difference is asymptotically vanishing (maybe in an oscillatory
manner) for large values of the wavevector.  However, recent exact
results,\cite{holas,stringari} and associated Quantum Monte Carlo
(QMC) calculations\cite{ceperley,static,conti} show that this is not
the case: instead, the aforementioned difference approaches a
{\it{negative constant}} as $q \to \infty$. In this paper, we revisit
the DFT of freezing, using the correct liquid-state input. We examine
the performance of the perturbative second-order theory (SOT)
\cite{ry} and the nonperturbative modified weighted density
approximation (MWDA).\cite{mwda} For a variety of systems, and
irrespective of the range of the interaction and the statistics
(superfluid {$^{4}\rm{He}$}, charged bosons and fermions) we find that
this {\it{exact}} large-$q$ behavior has drastic consequences in three
spatial dimensions: the crystal is predicted to be the stable phase
for any density. The SOT-functional is affected by this behavior most
dramatically: it becomes unbounded from below as the density becomes
more localized around the lattice sites and thus it has a
minus-infinity minimum at a perfectly localized density.  The MWDA, on
the other hand, does not suffer from this extreme pathology: the
MWDA-functional is bounded from below, but the (finite) minimum of the
energy always occurs for a modulated (crystal) phase. In two
dimensions, the effect is much less drastic, in the sense that for
densities relevant to crystallization the SOT-functional continues to
be bounded from below, yielding satisfactory predictions for the
freezing of the electron gas.

The rest of this paper is organized as follows: in Section II we
present a summary of the DFT formalism; in Section III we survey the
liquid-state input and discuss its implications on the behavior of the
`quantum' direct correlation functions, as well as on effective
interactions---in the electrons gas; in Section IV we apply the SOT
and in Section V the MWDA to the problem of freezing of different
quantum liquids. Finally, in Section VI we summarize and conclude.
\renewcommand{\thepage}{\arabic{page}}

\section{Quantum density functional theory of freezing}

The quantum DFT formalism employed in this work has been presented in
detail in Refs. \onlinecite{senatore,saverio}.  Here we give only an
outline and refer the reader to the above papers for details.  Writing
$E[n] = T_0[n]+E_{ex}[n]$ and using Eqs.  (\ref{var}) and
(\ref{equil}) we see that a {\it{necessary}} condition for equilibrium
is
\begin{eqnarray}
 \Biggl[{{\delta T_0[n]}\over{\delta n({\bf{r}})}} + {{\delta
   E_{ex}[n]}\over{\delta n({\bf{r}})}}\Biggr]_{n_0({\bf{r}})} = \mu -
   v_{ext}({\bf{r}}), \label{necess}
\end{eqnarray}
for the case of interacting particles. This is formally equivalent to
the condition of equilibrium for {\it{noninteracting}} particles (for
which $E[n] = T_0[n]$) under the influence of an effective external
potential
\begin{eqnarray}
   v_{eff}({\bf{r}}) = {{\delta E_{ex}[n]}\over{\delta n({\bf{r}})}}+
				   v_{ext}({\bf{r}}). \label{veff}
\end{eqnarray}
Note that the effective potential is itself a functional of the
one-particle density, through the dependence of $E_{ex}[n]$ on
$n({\bf{r}})$. Therefore, one is faced with a {\it{self-consistency}}
calculation which in practice proceeds as follows: an initial guess is
made for the density profile, which yields an initial form for the
effective potential.  Then the one-particle Schr\"odinger equations
(Kohn-Sham equations)
\begin{eqnarray}
   \Bigl[-{{\hbar^2}\over{2m}}\nabla^2 +
	v_{eff}({\bf{r}})\Bigr] \psi_i({\bf{r}}) =
   \varepsilon_i \psi({\bf{r}}) \label{shchroed} 
\end{eqnarray}
are solved, yielding the eigenfunctions $\psi_i({\bf{r}})$ and the
associated energy eigenvalues $\varepsilon_i$. From the former, a new
one-particle density is constructed through
\begin{eqnarray}
   n({\bf{r}}) = \sum_i n_i |\psi_i({\bf{r}})|^2, \label{dens} 
\end{eqnarray}
where $n_i$ are the occupation numbers suitable for the given
statistics (Bose or Fermi). The new density serves for the
construction of the new effective potential, and the cycle is
continued until a self-consistent solution has been found. Once the
self-consistent orbitals $\psi_i({\bf{r}})$ and the associated
eigenvalues $\varepsilon_i$ and density $n_0({\bf{r}})$ are known, the
ideal kinetic energy $T_0$ is given by
\begin{eqnarray}
   T_0 &=& \sum_i n_i \int d{\bf{r}} \psi_i^{*}({\bf{r}})
	\Bigl(-{{\hbar^2}\over{2m}}\nabla^2\Bigr) \psi_i({\bf{r}}) 
       \cr &=&
	\sum_i n_i \varepsilon_i - \int d{\bf{r}} n_0({\bf{r}})
	v_{eff}({\bf{r}}). \label{t0} 
\end{eqnarray}
The formulation presented above is {\it{exact}}, provided
$v$-representability holds.\cite{py89} Approximations enter through
the excess energy functional $E_{ex}[n]$ which is not known
exactly. In the following subsections we present two common schemes
which both rely on the knowledge of the second functional derivative
of this functional with respect to the density at the uniform limit.
This quantity is in turn directly related to the density-density
linear response function.

\subsection{Second-order theory}

One usual approximation is the so-called second-order theory (SOT) or
quadratic approximation. Here, one expands functionally the unknown
functional about a uniform fluid of density $n_l$, keeping terms up to
second-order only. Explicitly,
\begin{eqnarray}
  E_{ex}[n] &=& E_{ex}(n_l) + \int d{\bf{r}} {{\delta E_{ex}[n]}\over
  {\delta n({\bf{r}})}}\Bigg|_{n_l} \delta n({\bf{r}}) \cr &&
  +{1\over{2}} \int\int d{\bf{r}} d{\bf{r'}} {{\delta^2
  E_{ex}[n]}\over {\delta n({\bf{r}}) \delta
  n({\bf{r'}})}}\Bigg|_{n_l} \delta n({\bf{r}})\delta n({\bf{r'}}),
  \label{sot}
\end{eqnarray}
with $\delta n({\bf{r}})=n({\bf{r}})-n_l$ and $E_{ex}(n_l)$ the excess
intrinsic energy of the uniform liquid, a {\it{function}} of $n_l$.
Due to the translational and rotational invariance of the liquid, the
first functional derivative in the rhs of Eq. (\ref{sot}) is just a
position-independent constant, equal to the excess chemical potential
of the homogeneous liquid.  The second functional derivative is a
function of $|{\bf{r}}-{\bf{r'}}|$ only; both depend on $n_l$, of
course.  We define, from now on,
\begin{eqnarray}
    {{\delta^2 E_{ex}[n]}\over
    {\delta n({\bf{r}}) \delta n({\bf{r'}})}}\Bigg|_{n_l} \equiv
    -K(|{\bf{r}}-{\bf{r'}}|;n_l). \label{sfd} 
\end{eqnarray}
The function $K(r;n)$ is the excess part of the linear static inverse
response function of the homogeneous liquid, and can also be expressed
as \cite{senatore}
\begin{eqnarray}
   K(r;n) = \chi^{-1}(r;n) - \chi_0^{-1}(r;n), \label{chis} 
\end{eqnarray}
where $\chi^{-1}(r;n)$ and $\chi_0^{-1}(r;n)$ are the functional
inverses of the density-density static linear-response functions of
the interacting and noninteracting liquid, respectively.

Such an approximation is not {\it{a priori}} guaranteed to have any
validity, since there is no `small parameter' guiding the
expansion. Its widespread use is due on the one hand to practical
limitations, as third- and higher-order functional derivatives of
$E_{ex}[n]$ are poorly known even in the uniform phase, and on the
other hand in the relative success that it has had, at least in the
classical regime, in predicting the freezing parameters of simple
liquids.\cite{reviews} The function $K(r;n_l)$ is {\it{formally}} the
quantum analog of the classical Ornstein-Zernicke direct correlation
function (dcf).\cite{hansen}

We set $v_{ext}({\bf{r}})=0$ from now on. In the quadratic
approximation for the excess part of the energy functional, the
effective potential which enters in the Kohn-Sham calculation is
periodic with Fourier components
\begin{eqnarray}
  v_{eff}({\bf{Q}}) = 
  \delta n_{\bf{Q}}[-\tilde\chi^{-1}(Q;n_l)+\tilde\chi_0^{-1}(Q;n_l)], 
  \label{veffq} 
\end{eqnarray}
where ${\bf{Q}}$ is a reciprocal lattice vector (RLV) of the given
lattice and $\delta n_{\bf{Q}}$ is the Fourier component of the
periodic function $\delta n({\bf{r}}) \equiv n({\bf{r}})-n_l$, and
$\tilde\chi^{-1}(q;n_l)$ is the Fourier transform of the function
$\chi^{-1}(r;n_l)$.

For systems of neutral particles, the choice of the density $n_l$ of
the reference liquid is arbitrary, although the usual choice is to
consider a liquid at the same chemical potential as the
solid. Moreover, for a Bose system at $T=0$ the kinetic energy of
independent particles vanishes in the uniform limit. Thus, the
difference between the grand potential\cite{explain} 
of the solid and the liquid
is:\cite{moroni}
\begin{eqnarray}
   \Delta\Omega[n] = T_0[n] &-& {1\over{2}} \int\int
   d{\bf{r}}d{\bf{r'}} K(|{\bf{r}}-{\bf{r'}}|;n_l) \delta n({\bf{r}})
   \delta n({\bf{r'}})\hspace{-2cm} \cr \cr = T_0[n] &-&{V\over{2}}
   \tilde K(0;n_l) (n_s-n_l)^2 \cr \cr &&~~~~~~~~ - {V\over{2}}
   \sum_{{\bf{Q}}\ne {\bf{0}}} |n_{\bf{Q}}|^2 \tilde K(Q;n_l).
   \label{dom}
\end{eqnarray}
In Eq. (\ref{dom}), $V$ is the volume of the system, $n_s$ is the
average density of the solid, $\tilde K(q;n)$ denotes the Fourier
transform of $K(r;n)$ at wavevector $q$, and $n_{\bf{Q}}$ is the
Fourier component of the periodic density at RLV ${\bf{Q}}$.  In
practice, one changes $\mu$ (or, equivalently, $n_l$) and minimizes
$\Delta\Omega[n]$ with respect to $n({\bf r})$. Freezing occurs when
$\min\{\Delta\Omega[n]\}$ vanishes. For 
$\min\{\Delta\Omega[n]\} > 0\; (< 0)$ the
liquid (solid) is stable.

For systems composed of particles carrying a charge $e$ and
interacting via the Coulomb potential $v_c(r) = e^2/r$, the presence
of a uniform, rigid, neutralizing background of opposite charge
guarantees the stability of the system. The presence of the background
imposes the constraint that the freezing transition now takes place
{\it{at constant density}} (isochoric freezing). The relevant
thermodynamic potential is now the total energy $E[n]$; the phase with
the lowest $E[n]$ is the thermodynamically stable one. It is customary
for such systems to separate the excess energy into a Hartree
contribution and an `exchange-correlation' contribution, i.e. to write
\begin{eqnarray}
   E_{ex}[n] = {{e^2}\over{2}} \int\int d{\bf{r}} d{\bf{r'}} {{\delta
	n({\bf{r}}) \delta n({\bf{r'}})} \over{|{\bf{r}}-{\bf{r'}}|}}+
	E_{xc}[n], \label{xc}
\end{eqnarray}
where $\delta n({\bf{r}}) = n({\bf{r}}) - \bar n$ and $\bar n$ is the
average density. If we now define
\begin{eqnarray}
  {{\delta^2 E_{xc}[n]}\over {\delta n({\bf{r}}) \delta n({\bf{r'}})}}
 \Bigg|_{n_l} \equiv -K_{xc}(|{\bf{r}}-{\bf{r'}}|;n_l), \label{kxc}
\end{eqnarray}
then Eqs. (\ref{sfd}) and (\ref{xc}) imply
\begin{eqnarray}
   K(|{\bf{r}}-{\bf{r'}}|;n_l) = -v_c(|{\bf{r}}-{\bf{r'}}|) 
   +K_{xc}(|{\bf{r}}-{\bf{r'}}|;n_l). \label{sep} 
\end{eqnarray}
In Fourier space, one writes the Fourier transform $\tilde
K_{xc}(q;n)$ of $K_{xc}(r;n)$ as $\tilde K_{xc}(q;n) = v_c(q) G(q;n)$,
where $v_c(q)$ is the Fourier transform of the Coulomb potential
($v_c(q) = 4\pi e^2/q^2$ in three dimensions and $2\pi e^2/q$ in two
dimensions) and $G(q;n)$ is the so-called {\it{local field
factor}}.\cite{local} Finally we have
\begin{eqnarray}
   -\tilde K(q;n) = v_c(q)[1-G(q;n)]. \label{product} 
\end{eqnarray}

Due to the long-range nature of the Coulomb potential, the functional
expansion of the energy of the inhomogeneous phase can now be
performed only about a liquid whose density $n_l$ is {\it{equal}} to
the average density $\bar n \equiv n_s$ of the solid.  Using
Eqs. (\ref{sot}), (\ref{sfd}), (\ref{sep}) and (\ref{product}) we
obtain the difference between the energy\cite{explain} 
of the solid and the liquid
phases as:
\begin{eqnarray}
   \Delta E[n] &=& T_0[n] - {{d}\over{d+2}}N\epsilon_F \cr \cr && +
   {{V}\over{2}} \sum_{{\bf{Q}}\ne{\bf{0}}}|n_{\bf{Q}}|^2v_c(Q)
   [1-G(Q;n_s)].  \label{den}
\end{eqnarray}
Eq. (\ref{den}) above is valid for fermions in $d$-dimensions with
$\epsilon_F$ being the Fermi energy of noninteracting particles in the
liquid phase.  For bosons, this equation remains valid with the
omission of the second term in the rhs.

As mentioned above, the lack of a small parameter in the functional
expansion of the excess energy (at least as far as the freezing
problem is concerned) has cast some doubt on the validity of the
quadratic theory. This observation has led to the development of a
class of nonperturbative approximations, which approximate the excess
energy of the solid by that of a liquid. The density of the latter is
a weighted average of the true density of the solid.  Of particular
interest, due to its computational simplicity and its success in
describing bulk freezing for certain model systems in the classical
regime, is the modified weighted density approximation (MWDA) of
Denton and Ashcroft,\cite{mwda} presented in the following subsection.

\subsection{Modified weighted density approximation}

The MWDA amounts to the approximation of the excess energy of the
modulated system by that of a uniform system at a weighted density
\cite{mwda,saverio} with the latter being evaluated as a weighted
average over the real density of the crystal in a self-consistent
way. In other words, one writes
\begin{eqnarray}
   E_{ex}[n] \approx E^{MWDA}_{ex}[n] =
   N \epsilon (\hat n), \label{mw1}
\end{eqnarray}
where $\epsilon(n)$ is the excess energy per particle of a uniform
liquid of density $n$. The effective density $\hat n$ is evaluated as
a weighted average over the spatially-varying density $n({\bf{r}})$ of
the crystal and is defined by
\begin{eqnarray}
	\hat n = {{1}\over{N}}\int\int d{\bf{r}} d{\bf{r'}}
	n({\bf{r}}) n({\bf{r'}}) w({\bf{r}}-{\bf{r'}};\hat n),
	\label{defmwda}
\end{eqnarray}
where the weight function $w({\bf{r}}-{\bf{r'}};\hat n)$, which
depends on the weighted density itself is determined by requiring that
the MWDA functional is exact to second-order in a functional expansion
around a uniform liquid.  The derivation of the expression for the
weight function has been presented in detail elsewhere
\cite{saverio,mwda} and so here we show only the final results which
read as
\begin{eqnarray}
	w(r;\hat{n})= -\frac{1}{2 \epsilon'(\hat{n})}\left(K(r;
	\hat{n}) +\frac{\hat{n}}{V}\epsilon''(\hat{n})\right)
\label{weight}
\end{eqnarray}
and
\begin{eqnarray}
	\hat n = n_s + {{1}\over{n_s}} \sum_{{\bf{Q}}\ne {\bf{0}}}
	|n_{\bf{Q}}|^2 {{[-\tilde K(Q;\hat n)]}\over{2 \epsilon'(\hat
	n)}}.  \label{mw2}
\end{eqnarray}
The effective potential for the MWDA is readily calculated as
\cite{saverio}
\begin{eqnarray}
	v_{eff}({\bf{r}}) = \epsilon(\hat n) + \epsilon'(\hat n)
	{{\delta \hat n}\over{\delta n({\bf{r}})}}, \label{veffmwda}
\end{eqnarray}
and the corresponding expression in Fourier space, which is necessary
for the solution of the MWDA-Kohn-Sham equations can be found in
Ref. \onlinecite{saverio}.  The MWDA excess energy functional is exact
to second order in a functional expansion about a reference liquid,
but also includes contributions from all higher orders. In this sense,
the MWDA is a nonperturbative approximate scheme for the calculation
of the excess part of the energy.

\subsection{The Gaussian ansatz}
\label{gaussian}
The self-consistent solution of the Kohn-Sham equations is sometimes
avoided by taking advantage of the fact that, in the solid phase, the
particles are well localized around the lattice sites. This leads to
the introduction of the following Gaussian ansatz.  One constructs
normalized Bloch orbitals $\psi_{\bf{k}}({\bf{r}})$ from a single
Gaussian per site, $\phi(r) = (2\alpha/\pi)^{d/4}e^{-\alpha r^2}$,
according to:\cite{deber}
\begin{eqnarray}
   \psi_{\bf{k}}({\bf{r}}) = {{1}\over{\sqrt{N P_0({\bf{k}})}}}
   \Bigl({{2\alpha}\over{\pi}}\Bigr)^{d/4} \sum_{\bf{R}} e^{i
   {\bf{k}}\cdot{\bf{R}}} e^{-\alpha ({\bf{r}}-{\bf{R}})^2},
   \label{gauss}
\end{eqnarray}
where $\{{\bf{R}}\}$ is the set of Bravais lattice vectors and
\begin{eqnarray}
   P_m({\bf{k}}) = \sum_{\bf{R}} R^m e^{i {\bf{k}}\cdot{\bf{R}}-\alpha
   R^2/2}. \label{pm}
\end{eqnarray}
After some algebra, we arrive at the following explicit expressions
for the noninteracting kinetic energy $T_0$ and the Fourier component
of the density $n_{\bf{Q}}$:
\begin{eqnarray}
   T_0[n] = N {{\hbar^2}\over{2 m}}[d\alpha - \alpha^2 \mu_2]
   \label{fkea}
\end{eqnarray}
and
\begin{eqnarray}
   n_{\bf{Q}} = n_s e^{-Q^2/8\alpha} \mu_{\bf{Q}}, \label{fnuq}
\end{eqnarray}
where 
\begin{eqnarray}
   \mu_2 = {{\sigma}\over{N}}\sum_{\bf{k}}{{P_2({\bf{k}})}
   \over{P_0({\bf{k}})}}, \quad \mu_{\bf{Q}} =
   {{\sigma}\over{N}}\sum_{\bf{k}}{{P_0({\bf{k}}-{\bf{Q}}/2)}\over
   {P_0({\bf{k}})}}. \label{fmus}
\end{eqnarray}
Eqs. (\ref{fkea})-(\ref{fmus}) above, are valid for fermions; $\sigma$
denotes the number of particles in each occupied orbital: $\sigma = 1$
for spin-polarized and $\sigma = 2$ for unpolarized particles. The
${\bf{k}}$-sums extend over the occupied orbitals only. For bosons, we
have to put all the particles in the same orbital, ${\bf{k}} = 0$. In
this case Eqs.  (\ref{fkea}) and (\ref{fnuq}) remain valid with the
identification:
\begin{eqnarray}
   \mu_2 = {{P_2(0)}\over{P_0(0)}}, \qquad
   \mu_{\bf{Q}} = {{P_0({\bf{Q}}/2)}\over{P_0(0)}}. \label{bmus} 
\end{eqnarray}
Substituting the appropriate expression for $T_0$ and $n_{\bf{Q}}$
into Eqs. (\ref{dom}) or (\ref{den}) above, one directly obtains the
difference of the appropriate thermodynamic potential between the
solid and the liquid, within the SOT.  In the MWDA, the additional
self-consistent solution of Eq. (\ref{mw2}) is required to get the
excess energy of Eq. (\ref{mw1}).  In both cases one ends up with
differences of thermodynamic potentials as a function of $\alpha$,
$n_s$ and $n_l$. One then varies $\alpha$ (and $n_s$ for neutral
particles) until a minimum is found. By repeating the procedure for
different values of $n_l$ one can determine the phase diagram of the
system at hand.

We are going to present results obtained mainly through the use of the
Gaussian ansatz, rather than the full self-consistent calculation. The
reason is that, if a minimum exists when the Gaussian ansatz is
employed, then the full calculation can only yield a lower minimum
since the class of Gaussian densities is only a subclass of all the
possible profiles. Since our calculations yield {\it{too low minima}},
the self-consistent calculation is for most purposes
redundant. Moreover, the Gaussian ansatz allows for analytical
estimates of the magnitudes of the ideal and excess terms in
Eq. (\ref{dom}) or (\ref{den}).

The excess liquid-state linear static inverse-response function
$K(r;n)$ plays, evidently, a central role in the implementation of the
approximate schemes presented above. In the following Section we
discuss the form and asymptotic behavior of this function.

\section{Liquid-state input, quantum direct correlation function, and
effective interactions}
\label{dcfq}

For classical liquids, the Fourier transform of the dcf is related to
the experimentally measured structure factor $S(q)$ by a simple
algebraic relation,\cite{hansen} by virtue of the
fluctuation-dissipation theorem.  For quantum systems, on the other
hand, the theorem relates dynamical quantities, and the relation
between static quantities is not simple any more.\cite{senatore} As a
result, various approximations for the static linear response function
$\tilde\chi(q)$ have been developed.

Superfluid {$^{4}\rm{He}$} is a test case. This is a fluid of neutral
particles whose interactions can be accurately described by the
so-called Aziz potential.\cite{aziz,kalos} In the absence of accurate
data for $\tilde\chi(q)$, one often resorts to the Feynman
approximation to obtain a relation between $\tilde\chi(q)$ and
$S(q)$,\cite{moroni} which reads as
\begin{eqnarray}
  \tilde\chi_F(q;n_l) = \tilde\chi_0(q;n_l) S^2(q),
  \label{feynm1}
\end{eqnarray}
where $\tilde\chi_0(q;n_l) = -4mn_l/\hbar^2 q^2$ is the static
susceptibility of the ideal boson gas.  The ensuing approximate dcf
$\tilde K_F(q;n_l)$ has been employed in density-functional theories
of freezing of {$^{4}\rm{He}$} (Refs. \onlinecite{moroni,dalfovo}) or
Bose hard spheres \cite{denton} albeit with an appropriate `rescaling'
which was employed in an empirical way. This rescaling has been
avoided in a recent density-functional study of quantum hard-sphere
freezing.\cite{preprint} However, accurate data for $\tilde\chi(q)$
have now been obtained from diffusion Monte Carlo
calculations.\cite{ceperley}
\begin{figure}
\null
\vspace{-3.1cm}
\hspace{0cm}\psfig{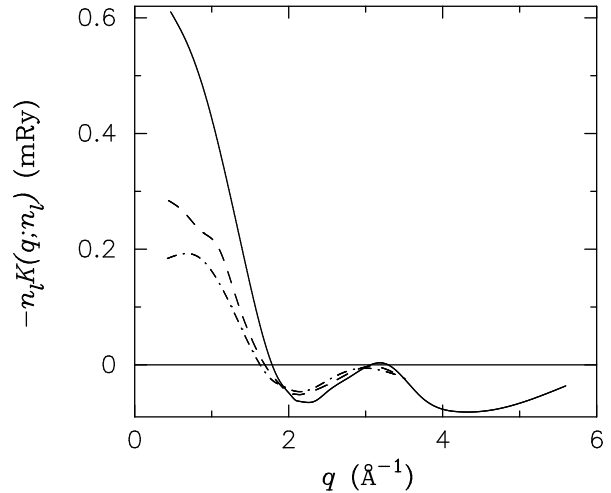}
\vspace{1.7cm}
\caption[dum01] {The function $-n_l\tilde K(q;n_l)$ (in mRy) of
superfluid {$^{4}\rm{He}$} as obtained from
simulations,\cite{ceperley} for three different fluid densities.
Solid line: $n_l = 0.02622$ {\AA}$^{-3}$; dashed line: $n_l = 0.02186$
{\AA}$^{-3}$; dash-dotted line: $n_l = 0.01964$ {\AA}$^{-3}$.}
\label{kofqhe}
\end{figure}
In Fig.~\ref{kofqhe} we show plots of this {\it{accurate}} direct
correlation function for three different densities of the liquid. A
comparison with the Feynman approximation \cite{moroni} shows
immediately that whereas the latter has an oscillatory behavior about
zero, the exact dcf is negative for almost all values of $q > 2$
{\AA}$^{-1}$.  An additional important difference concerns the
large-$q$ behavior of the dcf. Although the Monte Carlo data are
limited to values $ q < 4-6$ {\AA}$^{-1}$, exact theoretical
calculations show that the $q \to \infty$-limit of $-\tilde K(q,n_l)$
is a negative number, and not zero as the Feynman approximation
implies.\cite{stringari} In particular, the response function
$\tilde\chi(q;n_l)$ is given for large $q$ by: \cite{stringari}
\begin{eqnarray}
    \tilde\chi(q;n_l) = -{{4mn_l}\over{\hbar^2 q^2}} \Bigl[ 1 +
   {{8m}\over{3 \hbar^2 q^2}}\langle KE \rangle + O(q^{-4})\Bigr],
   \label{string}
\end{eqnarray}
From Eqs. (\ref{chis}), (\ref{string}) and using the result
$\tilde\chi_0(q;n_l)=-4mn_l/\hbar^2 q^2$ for the static susceptibility
of the ideal boson gas, we obtain
\begin{eqnarray}
   -n_l \tilde K(\infty;n_l) = -{{2}\over{3}} \langle KE \rangle,
   \label{khe}
\end{eqnarray}
where $\langle KE \rangle$ is the expectation value of the kinetic
energy in the liquid phase.  These features of the exact dcf have
important consequences on the performance of DFT's of freezing, as
will be shown below.

Charged fermions or bosons are another example of quantum liquids.
The former is just the usual system of electrons in a uniform
background (jellium) and the latter is a model system of spinless
particles of electronic charge $e$ and mass $m$ in a background, but
obeying Bose statistics. A natural length scale for these systems is
the so-called Wigner-Seitz radius $r_0$ defined as the radius of a
sphere which contains, on average, one particle, i.e. for a system of
density $n$ in $d$-dimensions we have
\newpage
\begin{figure}
\null
\vspace{-3.1cm}
\hspace{0cm}\psfig{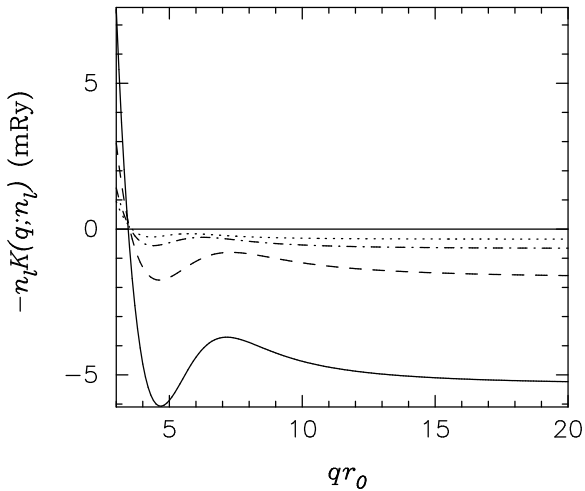}
\vspace{2cm}
\caption[dum02] {The function $-n_l\tilde K(q;n_l)$ (in mRy) of
charged bosons as obtained from simulations:\cite{conti} solid line:
$r_s = 20$; dashed line: $r_s = 50$; dash-dotted line: $r_s = 100$;
dotted line: $r_s = 160$.}
\label{kofqch}
\end{figure}
\begin{eqnarray}
 n = {{3}\over{4\pi r_0^3}} \;\; {\rm(} d=3 {\rm)}, \qquad n =
 {{1}\over{\pi r_0^2}} \;\; {\rm(} d=2 {\rm)}. \label{ws}
\end{eqnarray} 
A convenient dimensionless measure of the density is $r_s\equiv
r_0/a_0$, where $a_0$ is the Bohr radius.  A widely used scheme to
relate the local field factor $G(q)$ with the structure factor has
been introduced by Singwi {\it{et al.}} \cite{stls} and is denoted by
STLS. This has been employed in DFT's of freezing of jellium in a
number of cases.\cite{senatore,saverio} An important feature of the
STLS scheme is that in the limit of large-$q$ the local field factor
$G(q)$ approaches unity and this implies that $-\tilde K(q;n_l)$
approaches zero in that limit [see Eq. (\ref{product})]. In this
respect, the STLS scheme for systems of charged particles has the same
features as the Feynman approximation. However, it has been shown
{\it{exactly}} that in the large-$q$ limit, $G(q)$ goes like $q^2$ in
three dimensions; \cite{holas,stringari} moreover it can easily be
shown that it scales like $q$ in two dimensions.\cite{rochester} In
particular, for charged bosons at $d=3$ it is known that,\cite{conti}
for large $q$,
\begin{eqnarray}
     G(q;n_l) &=& \frac{2 \langle KE \rangle q^2}{3 m \omega_{pl}^2}+
 {2\over{3}}(1-g(0)) + \frac{16\langle (KE)^2 \rangle}{ 5\hbar^2
 \omega_{pl}^2} \cr \cr && - \frac{16\langle KE \rangle^2}{ 9\hbar^2
 \omega_{pl}^2} +O(q^{-2}), \label{gbos}
\end{eqnarray}
where $\omega_{pl} = \sqrt{4\pi n_l e^2/m}$ is the plasma frequency
and $g(0)$ is the value of the pair distribution function of the
liquid $g(r)$ at zero separation. From Eqs.  (\ref{product}) and
(\ref{gbos}) we find once more
\begin{eqnarray}
    -n_l \tilde K(\infty;n_l)= -{{2}\over{3}} \langle KE \rangle,
	\label{kbos}
\end{eqnarray}
as in Eq. (\ref{khe}) above.

For fermions in three dimensions, the large $q$ local field factor
reads as \cite{holas}
\begin{eqnarray}
  G(q;n_l) &=& \frac{2 (\langle KE \rangle-\langle KE \rangle_0) q^2}
{3 m \omega_{pl}^2} + {2\over{3}}(1-g(0)) \cr \cr && + {{16 (\langle
(KE)^2 \rangle -\langle (KE)^2 \rangle_0)}\over{5 \hbar^2
\omega_{pl}^2}} \cr\cr && - {{16 {(\langle KE \rangle^2 -\langle KE
\rangle_0^2)}}\over{9 \hbar^2 \omega_{pl}^2}} +O(q^{-2}),
\label{gfer}
\end{eqnarray}
where $\langle \cdots \rangle_0 $ denotes a noninteracting average,
and the coefficient of the $q^2$ term---the difference in the kinetic
energy per particle between the interacting and the noninteracting
system---is a positive quantity.\cite{holas} Note that the differences
between Eq. (\ref{gfer}) and Eq. (\ref{gbos}) arise from the different
momentum distributions of the noninteracting Fermi and Bose systems.
Using Eqs.  (\ref{product}), (\ref{gfer}) we finally obtain,
\begin{eqnarray}
    -n_l \tilde K(\infty;n_l)= -{{2}\over{3}} (\langle KE
\rangle-\langle KE \rangle_0).
\end{eqnarray}

\begin{figure}
\null
\vspace{-3.1cm}
\hspace{0cm}\psfig{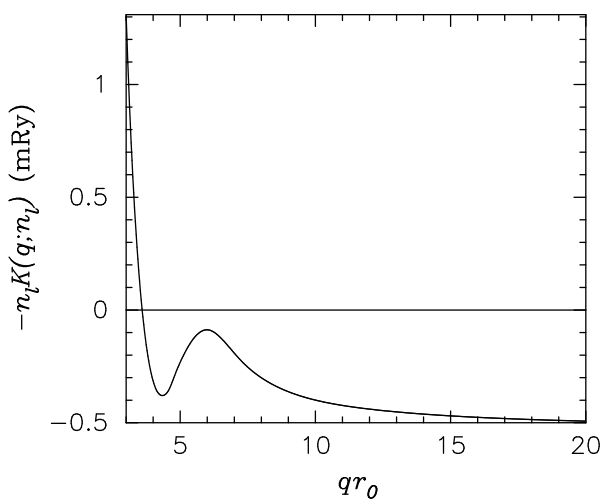}
\vspace{1.8cm}
\caption[dum02] {The function $-n_l\tilde K(q;n_l)$ (in mRy) of
spin-polarized charged fermions, as obtained from simulations,
\cite{review} at $r_s = 100$.\hfill\hfill}
\label{kofqch1}
\end{figure}

In Fig.~\ref{kofqch} we show the direct correlation function of
charged bosons for a number of different densities as obtained from
Quantum Monte Carlo simulations.\cite{conti} In Fig.~\ref{kofqch1} we
show the same function for fully polarized charged fermions at
$r_s=100$, which has also been obtained from QMC.\cite{review} In both
cases, it is clearly seen that at large values of $q$ the function
$-\tilde K(q;n_l)$ tends to a negative constant.  In the system of
point charged particles, by virtue of the virial theorem, this
constant may be expressed most simply as
\begin{eqnarray}
-n_l \tilde K(\infty;n_l) =\frac{2}{3}\frac{d(r_sE)}{dr_s},
\end{eqnarray}
with $E=\epsilon_c(r_s)$ the correlation energy per particle, for
fermions, and $E=\epsilon(r_s)$ the energy per particle, for bosons.

\begin{figure}
\null
\vspace{-3.3cm}
\hspace{0cm}\psfig{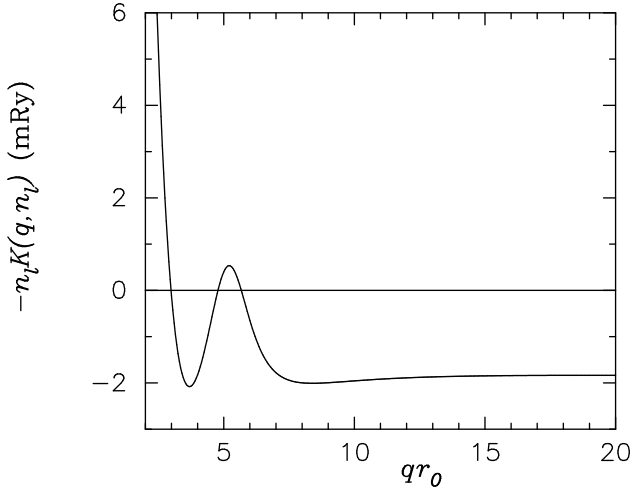}
\vspace{1.3cm}
\caption[dum021]
{The function $-n_l\tilde K(q;n_l)$ (in mRy)
of spin-polarized charged fermions in $2d$, as obtained
from simulations \cite{review} at $r_s=40$.}
\label{kofqch2}
\end{figure}

In two dimensions, the situation is quite similar. For fermions, using
the asymptotic behavior of the static linear response
function,\cite{yagar} it has been shown \cite{rochester} that the
local field factor scales linearly with $q$, as $q\to\infty$, namely
\begin{eqnarray}
   G(q;n_l) = \frac{(\langle KE \rangle-\langle KE \rangle_0)}{2\pi
e^2 n_l} q + 1 - g(0) +O(q^{-1}).\hspace{-1cm} \cr \label{twod}
\end{eqnarray}
From Eqs. (\ref{product}) and (\ref{twod}) we obtain
\begin{eqnarray} 
-n_l \tilde K(\infty;n_l) = -(\langle KE \rangle-\langle KE
\rangle_0).
\label{ktwod}
\end{eqnarray}

In Fig. ~\ref{kofqch2} we show the direct correlation function of
fully polarized electrons in 2 dimensions, near freezing, i.e., at
$r_s=40$, as obtained from Quantum Monte Carlo
simulations.\cite{review} Again, the saturation of $\tilde K(q,n_l)$
to a constant---which may be conveniently expressed as
\begin{eqnarray}
-n_l \tilde K(\infty;n_l) = -(\langle KE \rangle-\langle KE
\rangle_0)=\frac{d(r_s\epsilon_c)}{dr_s},
\label{k2limit}
\end{eqnarray}
with $\epsilon_c(r_s)$ the correlation energy per particle---is
evident. We note that the large $q$ behavior of $-n_l\tilde K(q,n_l)$
for all the systems considered above, is given by
\begin{eqnarray}
    -n_l \tilde K(q,n_l)&=& -{{2}\over{d}} (\langle KE \rangle -
	\langle KE \rangle_0) \cr\cr &&+ O(q^{-d+1})
	+O(q^{-2}),\hspace{-0.7cm}
\label{kgen}
\end{eqnarray}
and evidently for the noninteracting Bose systems $\langle KE
\rangle_0 =0$. In fact one may easily
show\cite{vignale,stringari,preprint} that Eq. (\ref{kgen}) above is
valid for any quantum liquid interacting with pair potentials, both in
in 3 and 2 dimensions, provided the second term on the rhs is only
retained for Coulombic systems ($1/r$ interaction) in 2 dimensions.

The short-wavelength behavior of $-\tilde K(q;n_l)$ described above,
implies that in real space the function $-K(r;n_l)$ has a
delta-function contribution at the origin with negative weight, as is
clear from Eq. (\ref{kgen}). We shall therefore define a regular dcf
$\tilde K_R(q;n_l)$, decaying to zero as $q\to\infty$, by setting
\begin{eqnarray}
    \tilde K(q;n_l) = \tilde K_R(q;n_l) + \frac{2}{ n_l d} (\langle KE
	\rangle - \langle KE \rangle_0).
\label{regular_q}
\end{eqnarray}
This implies in real space
\begin{eqnarray}
    K(r;n_l) &=& K_R(r;n_l) + U_0(n_l) \delta({\bf r})n_l^{-1}\cr\cr
             &\equiv& K_R(r;n_l) + K_S(r;n_l),
\label{regular_r}
\end{eqnarray}
with the strength of the singular part $K_S(r;n_l)$ given by
\begin{eqnarray}
    U_0(n_l) = \frac{2}{d}(\langle KE \rangle - \langle KE \rangle_0
    )>0.
\label{u0}
\end{eqnarray}
Before investigating the consequences of this unexpected behavior of
$-K(r;n_l)$ on the density functional theories of freezing, we shall
pause here to briefly discuss its implications on effective
interparticle interactions in the liquid phase. As an example we shall
consider spin unpolarized electrons in 3 dimensions.

Within the dielectric formalism, the number and spin linear response
functions of the normal electron fluid may be cast in a mean field,
RPA--like form, by defining appropriate polarization
potentials.\cite{pines85} In the static limit to which we shall
restrict here, the number and spin response functions read
respectively
\begin{eqnarray}
\tilde{\chi}(q)= \frac{\tilde{\chi}_0(q)}{1-V^s(q)\tilde{\chi}_0(q)}
	\label{chinum}
\end{eqnarray}
and 
\begin{eqnarray}
\tilde{\chi}_s(q)=
	-\mu_B^2\frac{\tilde{\chi}_0(q)}{1-V^a(q)\tilde{\chi}_0(q)},
	\label{chispin}
\end{eqnarray}
with $\mu_B$ the Bohr magneton and $V^s(q)$ and $V^a(q)$ the symmetric
and asymmetric polarization potentials,
respectively.\cite{pines85,yagar} From Eqs. (\ref{chis}) and
(\ref{product}) it follows that
\begin{eqnarray}
V^s(q)=-\tilde{K}(q,n_l)= v_c(q)[1-G^s(q,n_l)],
\label{Gsy}
\end{eqnarray}
with $G^s(q,n_l)\equiv G(q,n_l)$.  In a similar fashion one can
set\cite{yagar}
\begin{eqnarray}
V^a(q)= -v_c(q)G^a(q,n_l),
\label{Gasy}
\end{eqnarray}
which defines the asymmetric local field factor $G^a(q,n_l)$, whose
behavior for large $q$ is easily obtained from the known asymptotic
expansions of $\tilde{\chi}_s(q)$\cite{yagar} as 
\begin{eqnarray}
G^a(q,n_l) = G^s(q,n_l) -1 +2g(0) +O(q^{-2}). \label{gasyq}
\end{eqnarray}

Interparticle polarization potentials for pairs of electrons with
parallel or antiparallel spin projections are readily obtained from
their symmetric and asymmetric counterparts $V^s(q)$ and $V^a(q)$
via\cite{pines85}
\begin{eqnarray}
V_{\sigma\sigma'}^{\rm pol}(q)&=&V^s(q)\pm
V^a(q)\cr&=&v_c[1-G^s(q,n_l)\mp G^a(q,n_l)],
\label{pot}
\end{eqnarray}
where the upper sign corresponds to $\sigma\sigma'=\uparrow\uparrow$
and the lower to $\sigma\sigma'=\uparrow\downarrow$.  For large $q$,
from Eqs. (\ref{gfer}) and (\ref{gasyq}) one obtains
\begin{eqnarray}
V_{\uparrow\uparrow}^{\rm pol}(q)= -\frac{4}{3n_l}(\langle KE \rangle
	- \langle KE \rangle_0) +O(q^{-2}).
\label{ppotq}
\end{eqnarray}
and 
\begin{eqnarray}
 V_{\uparrow\downarrow}^{\rm pol}(q)=2g(0)v_c(q) +O(q^{-4}).
\label{upotq}
\end{eqnarray}
Eq. (\ref{ppotq}) implies the presence in $V_{\uparrow\uparrow}^{\rm
pol}(r)$ of a term $-2U_0(n_l)\delta({\bf r})/n_l$, with $U_0(n_l)>0$,
given by Eq. (\ref{u0}) with $d=3$. On the other hand from
Eq. (\ref{upotq}) one obtains that, for $r\to 0$, $
V_{\uparrow\downarrow}^{\rm pol}(r)=2g(0)e^2/r$. This looks quite
strange at first, as one would naively expect that at short distance
effective interelectronic interactions should be essentially
Coulombic. In fact, polarization potentials are not effective
potentials, though at times this is not appreciated. We should also
mention that in the approach of Refs. \onlinecite{pines85,iwamoto84}
the polarization potentials were assumed regular at the origin,
$V_{\sigma\sigma'}^{\rm pol}(0)=e^2q_{\sigma\sigma'}$, with the
screening wavevectors $q_{\sigma\sigma'}$ of the order of the Fermi
wavevector $q_F$.

Effective electronic interactions have been defined for the electrons
gas by Kukkonen and Overhauser a long time ago,\cite{kukover} using
the polarization potential method but keeping into account particle
indistinguishability. According to this study effective two--body
electron-electron interactions  may be written as 
\begin{eqnarray}
V_{\sigma\sigma'}(q)= v_c(q)\left[1+\Delta_{\sigma\sigma'}(q)\right],
\end{eqnarray}
where
\begin{eqnarray}
\Delta_{\sigma\sigma'}(q)&=&\frac{v_c(q)[1-G^s(q)]^2\chi_0(q)}
{1-v_c(q)[1-G^s(q)]\chi_0(q)} \cr \cr &&\pm 
\frac{v_c(q)[G^a(q)]^2\chi_0(q)}
{1+v_c(q)G^a(q)\chi_0(q)},
\end{eqnarray}
with the upper (lower) sign corresponding to parallel (antiparallel)
spins. The large $q$ behavior of $\Delta_{\sigma\sigma'}(q)$ is easily
obtained from Eqs. (\ref{gfer}), (\ref{gasyq}), and from the known
asymptotic behavior\cite{yagar} of the Lindhard function
\begin{eqnarray}
    \tilde\chi_0(q;n_l)_{q\to\infty} = 
   -{{4mn_l}\over{\hbar^2 q^2}}. 
	\label{chilin} 
\end{eqnarray}
One finds that $\Delta_{\uparrow\downarrow}(q)$ vanishes as
$q^{-2}$ for $q\to\infty$, while  
\begin{eqnarray}
\Delta_{\uparrow\uparrow}(q)_{q\to\infty}= -\frac{8}{27}r_s^3
\left[\frac{d(r_s\tilde{\epsilon}_c)}{dr_s}\right]^2,
\end{eqnarray}
with $\tilde{\epsilon}_c(r_s)$ the correlation energy per particle, in
Rydbergs, of the electron gas.  Thus $V_{\uparrow\downarrow}(r)=e^2/r$
for small $r$ while the effective interaction between parallel spin is
very slightly reduced with respect to the bare Coulomb repulsion,
$V_{\uparrow\uparrow}(r)=\gamma(r_s)e^2/r$, with
$\gamma(r_s)=1+\Delta_{\uparrow\uparrow}(\infty)\alt 1$.  In
particular, in the metallic regime one obtains from the known equation
of state of the electron gas\cite{alder,vosko} $\gamma(r_s)=0.99$ and
$0.98$ for $r_s=2$ and $5$.  Thus, as we anticipated, effective
interactions do remain essentially Coulombic at short distances.

\section{Second-Order Theory}
\subsection{Three dimensions}

We begin with the application of the second-order theory (SOT) to the
freezing of superfluid {$^{4}\rm{He}$}. Experimental results on the
system show that {$^{4}\rm{He}$} crystallizes\cite{grilly} at a liquid
density $n_l = 0.0260$ \AA$^{-3}$ into an hcp-solid of density $n_s =
0.0287$ \AA$^{-3}$.  We employ the Gaussian ansatz for a fcc-crystal
density and apply Eq. (\ref{dom}) for the evaluation of the
grand-potential difference between the solid and the liquid. The value
of $\tilde K(q,n_l)$ at $q=0$ which enters in this calculation is
related to the energy per particle $\epsilon(n_l)$ of the liquid via
the `compressibility sum rule', namely
\begin{eqnarray}
   -\tilde K(0;n_l) = 2\epsilon'(n_l) + n_l\epsilon''(n_l), \label{compr} 
\end{eqnarray}
where the primes denote differentiation with respect to the argument.
For the quantity $\epsilon(n_l)$ we use an analytic fit based on
accurate diffusion Monte Carlo data.\cite{euler}

\begin{figure}
\null
\vspace{-3.cm}
\hspace{.3cm}\psfig{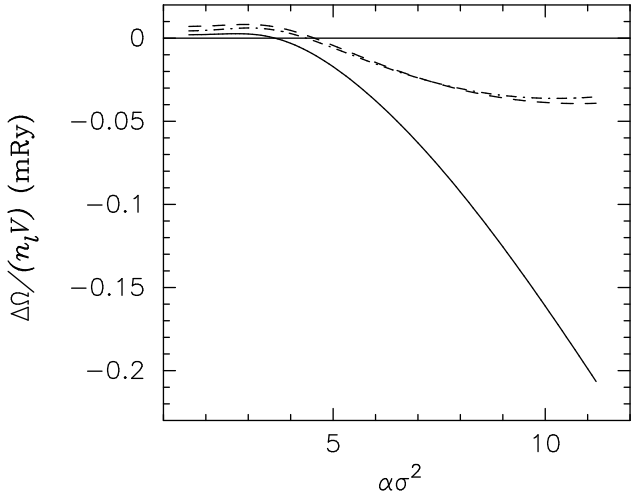}
\vspace{1.5cm}
\caption[dum03]  {Grand potential difference\cite{explain} [Eq. (\ref{dom})] 
between a 
$^4$He fcc-solid and a liquid at the same chemical potential, 
for different pairs
$(n_s,n_l)$ in the second-order theory.  Solid line: $n_s=0.0287$
\AA$^{-3}$, $n_l=0.0262$ \AA$^{-3}$; dashed line: $n_s=0.0287$
\AA$^{-3}$, $n_l=0.0216$ \AA$^{-3}$; dash-dotted line: $n_s=0.0275$
\AA$^{-3}$, $n_l=0.019$ \AA$^{-3}$.  Here, $\sigma = 2.556$ \AA.}
\label{domhe}
\end{figure}

We try to minimize $\Delta \Omega[n]$ [Eq. (\ref{dom})] with respect
to $\alpha$ for a variety of different values of $(n_s,n_l)$. As can
be seen in Fig.~\ref{domhe} for the pair of values which are close to
those for which freezing occurs in experiments, $\Delta\Omega[n]$ has
apparently {\it{no minimum}}; it keeps getting lower without bound as
the localization increases. In the same figure it can be seen that for
$n_l$ much lower than the freezing value, $\Delta\Omega$ has a very
negative local minimum at strong localization (large values of
$\alpha$), i.e. the solid is predicted to bee too stable. With
reference to Fig. ~\ref{domhe}, note that at freezing one would expect
$\alpha\sigma^2\approx 2$, in order to obtain the correct `quantum'
Lindemann ratio $\gamma\simeq 0.3$. ($\gamma$ is the ratio of the root
mean square deviation about a site to the nearest neighbor distance.)
On the contrary, the minima shown in the figure are at
$\alpha\sigma^2\approx 10$, implying a value of $\gamma \propto
1/\sqrt{\alpha}$ which is too small by about a factor 2, being
essentially classical. Unfortunately, the lack of Monte Carlo data for
the dcf at large wavevectors does not allow us to examine the limit of
strong localizations, since as $\alpha$ grows we need more and more
shells of RLVs into the sum of Eq. (\ref{dom}) in order to achieve
convergence. Nevertheless, it is clear from the shape of $-\tilde
K(q;n_l)$ (see Fig.~\ref{kofqhe}) that the overestimation of the
stability of the solid is brought about by the fact that $-\tilde
K(q,n_l)$ is {\it{negative}} for all values of the RLVs; this way, the
contribution from the excess part of the energy, which is becoming
lower with increasing localization, dominates over the contribution
from the ideal energy, which grows with localization, to yield a total
energy which is {\it{too low}}. We will make this statement more
quantitative shortly.

\begin{figure}
\null
\vspace{-3.1cm}
\hspace{.3cm}\psfig{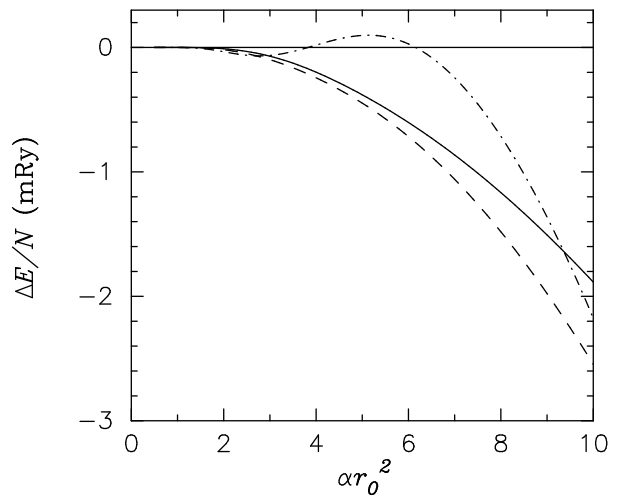}
\vspace{1.8cm}
\caption[dum04] {Ground-state energy difference\cite{explain}
[Eq. (\ref{den})] between a charged boson bcc-solid and the liquid,
versus localization at three different average densities, as obtained
from the second-order theory. Solid line: $r_s = 160$; dashed line:
$r_s = 100$; dash-dotted line: $r_s = 50$.}
\label{dench}
\end{figure}

Next, we look at the SOT-freezing of charged bosons, using the
dcf-simulation results of Ref. \onlinecite{conti}. The system is known
to undergo Wigner crystallization into a bcc-solid\cite{alder} at $r_s
= 160 \pm 10$.  Once more, we employ the Gaussian ansatz and try to
minimize $\Delta E[n]$ [Eq. (\ref{den})] with respect to $\alpha$ at
various different values of $r_s$. Some of the results are shown in
Fig.~\ref{dench}. The quantity $\Delta E[n]$ is clearly unbounded from
below, i.e. the absolute minimum lies at infinite localization, where
the value of $\Delta E[n]$ is {\it{minus infinity}}. There is a local
negative minimum at $\alpha r_0^2\approx 3$ for $r_s=50$. This
corresponds to the correct quantum Lindemann ratio, and one might
argue that the SOT of freezing may only make sense for modulations
that are not too large (i.e., moderate values of $\alpha$) and near
the freezing density. Even so, the predicted freezing density would be
overestimated by a factor of about $30$.

It becomes clear, therefore, that the SOT suffers from the pathology
of producing an unbounded functional. Within the framework of the
Gaussian approximation, this feature can be clearly understood as
follows. Take a solid whose lattice constant is $a$ and consider the
strong-localization limit, i.e. $\tilde\alpha \equiv \alpha a^2 \gg
1$.  In that case we have, with excellent accuracy, $\mu_2 = 0$ and
$\mu_{\bf{Q}} = 1$ for all ${\bf{Q}}$'s [see Eq. (\ref{bmus})]. In
$d$-dimensions, Eq. (\ref{fkea}) gives
\begin{eqnarray}
   {{T_0(\tilde\alpha)}\over{Vn_l}}\Bigg|_{\tilde\alpha \gg 1} =
   (1+\eta){{\hbar^2 d}\over{2ma^2}} \tilde \alpha, \label{t3d} 
\end{eqnarray}
where $n_s = (1+\eta)n_l$ ($\eta = 0$ for isochoric freezing).  On the
other hand, the excess energy contribution [i.e. the sum of the terms
beyond $T_0$ on the rhs of Eq. (\ref{dom}) or Eq. (\ref{den})] may be
conveniently broken into two contributions originating respectively
from $\tilde{K}_R(q; n_l)$ and $\tilde{K}_S(q; n_l)$. The first
contribution is most easily treated in reciprocal space, where it
takes the form
\begin{eqnarray}
   {{\Delta E_{ex}^R(\tilde\alpha)}\over{Vn_l}}\Bigg|_{\tilde\alpha
   \gg 1} &=& -{{\eta^2}\over{2}}n_l\tilde K_R(0;n_l) -
   {{(1+\eta)^2}\over{2}} \cr &&\times \sum_{{\bf{Q}}\ne 0}
   e^{-(Qa)^2/4\tilde\alpha} n_l \tilde K_R(Q;n_l), \hspace{-1cm}
\end{eqnarray}
and manifestly tends to a constant for large values of $\alpha$.  The
second contribution is evaluated in real space, to leading order, as
\begin{eqnarray}
   {{\Delta E_{ex}^S(\tilde\alpha)}\over{Vn_l}}\Bigg|_{\tilde\alpha
   \gg 1} &=& -\frac{U_0(n_l)}{2Vn_l^2}\int d{\bf r} (\delta n({\bf
   r}))^2 \cr &=&
   -\frac{U_0(n_l)(1+\eta)}{2n_la^d\pi^{d/2}}\tilde{\alpha}^{d/2} ,
\label{exca}  
\end{eqnarray}
using the fact that, for $\tilde{\alpha}\gg1$, the density reduces to
a superposition of nonoverlapping normalized Gaussians
$(2\alpha/\pi)^{d/2} \exp\{-2\alpha r^2\}$, one per site.  The lack of
a lower bound for $d=3$ can be now easily understood: since the ideal
energy scales like $\tilde \alpha$ and the excess like
$-\tilde\alpha^{3/2}$ for large $\tilde\alpha$, it is then clear that
their sum will be dominated by the $-\tilde\alpha^{3/2}$-term and will
be unbounded from below.  The analysis presented above is valid also
for fermions.

If a minimization within the restricted space of Gaussian profiles
fails to yield a finite minimum, then the {\it{global}} minimum of the
unrestricted Kohn-Sham (KS) scheme, which cannot be higher, will also
be minus infinity. This does not exclude, however, the possibility of
obtaining, by means of solving the KS-equations, some {\it{local}}
minimum at a moderate value of the localization; in the KS-scheme
there is no `localization parameter' of course, but the Lindemann
ratio, for example, can be used as a measure of the spatial extent of
the one-particle density around a lattice site. This possibility is
particularly interesting because it could be argued that the
perturbative character of the SOT immediately limits its validity to
weakly-modulated density profiles.  In order to pursue this line, we
have also performed the full, self-consistent
calculation\cite{senatore,saverio} for both charged bosons and
spin-polarized electrons, within the SOT, at the densities for which
the dcf is available (see Figs. \ref{kofqch} and \ref{kofqch1}
above). However, no local minimum was found, in either case. The
large-$q$ behavior of the dcf and the number of space dimensions
render the SOT {\it{pathological}} in $d=3$.

\begin{figure}
\null
\vspace{-2.8cm}
\hspace{.3cm}\psfig{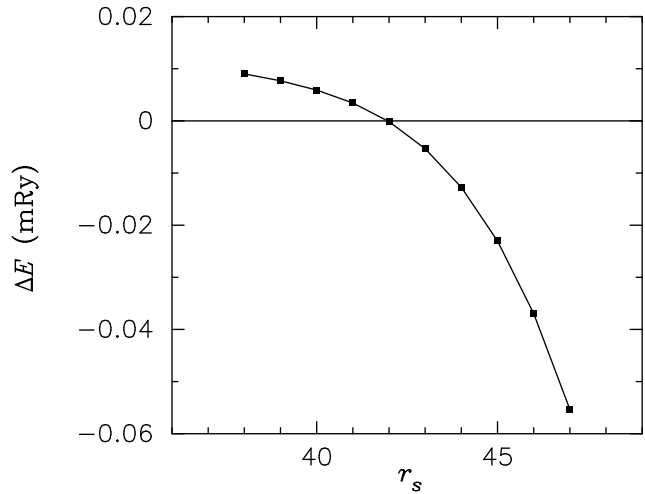}
\vspace{1.2cm}
\caption[dum04] {Ground state energy difference between
triangular-solid and polarized liquid for electrons in $2d$ as function
of the density. $\Delta E/N$ (in mRy) was calculated within the
SOT. The solid squares are calculated points, with the line just a
guide for the eye.}
\label{denche2}
\end{figure}

\subsection{Two dimensions}  

In two dimensions, Eqs. (\ref{t3d}) and (\ref{exca}) show that both
the ideal and excess term scale as $\tilde{\alpha}$ at strong
localizations, the former with a positive and the latter with a
negative coefficient.  Therefore, the absolute values of the
respective coefficients are crucial in determining the existence of a
lower bound for the sum of the two terms. Expressing energies in
Rydbergs and making use of Eqs.  (\ref{t3d}), (\ref{exca}),
(\ref{u0}), and (\ref{k2limit}) we find for isochoric transitions
($\eta = 0$):
\begin{eqnarray}
  {{T_0(\tilde\alpha;r_s)}\over{N}}\Bigg|_{\tilde\alpha \gg 1} = 2
  \tilde\alpha \Bigl({{a_0}\over{a}}\Bigr)^2
\end{eqnarray}
and
\begin{eqnarray}
  {{\Delta E_{ex}(\tilde\alpha;r_s)}\over{N}}\Bigg|_{\tilde\alpha \gg
  1} = \frac{1}{2}r_s^2\frac{d(r_s\tilde{\epsilon}_c)}{dr_s}
  \tilde\alpha \Bigl({{a_0}\over{a}}\Bigr)^2, \label{comp2d}
\end{eqnarray}
where $a$ is the lattice constant of the given crystal structure and
$\tilde{\epsilon}_c$ the correlation energy per particle in
Rydbergs. Thus, at strong localizations
\begin{eqnarray}
  {{\Delta E(\tilde\alpha;r_s)}\over{N}}\Bigg|_{\tilde\alpha \gg 1} \propto
  \Bigl(2 + \frac{1}{2}r_s^2\frac{d(r_s\tilde{\epsilon}_c)}{dr_s}
  \Bigr) \tilde\alpha.  \label{2dtotal}
\end{eqnarray}
For polarized fermions ($\sigma = 1$) in $2d$ the available QMC
data\cite{tanatar,rapisard} show that the coefficient in
(\ref{2dtotal}) is {\it{positive}} for $r_s \leq 59$. Therefore, the
SOT-functional remains bounded from below for values $r_s \leq
59$. This is encouraging, given that the polarized electron gas in two
dimensions crystallizes into a triangular lattice at a value of $r_s$
which is considerably smaller than this `stability limit'.

\begin{figure}
\null
\vspace{-2.9cm}
\hspace{.3cm}\psfig{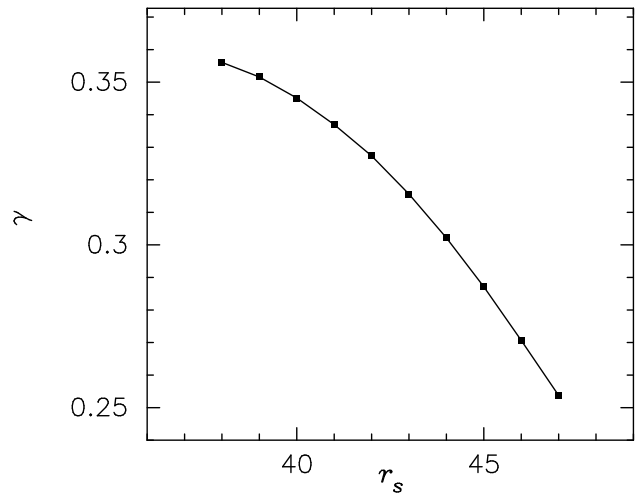}
\vspace{1.5cm}
\caption[dum04] {Lindemann ratio $\gamma$ around freezing in the
triangular $2d$ Wigner crystal, as predicted by the SOT. The solid
squares are calculated points, with the line just a guide for the
eye.}
\label{linde}
\end{figure}

We have thus performed a full Kohn-Sham calculation with the accurate
liquid state input shown in Fig. \ref{kofqch2}, using a plane wave
basis set as explained at length elsewhere.\cite{senatore,saverio} We
have systematically checked convergence with respect to the plane wave
cutoff and the number of $k$-points in the Brillouin zone. As it can
be seen in Fig. \ref{denche2} at $r_s=40$, where we have the dcf from
QMC,\cite{review} the solid is still unstable, though its energy is
only $6$ microRydbergs higher than that of the polarized liquid. On
the ground that the explicit dependence of $G(q/q_F; r_s)$ on $r_s$
should be very weak,\cite{review} we have neglected it altogether to
perform the calculations at the other values of $r_s$, using therefore
the available local field factor $G(q/q_F; r_s=40)$. This treatment
predicts freezing from the polarized fluid at $r_s=42$ which agrees
within error bars with the QMC prediction of Tanatar and
Ceperley\cite{tanatar} $r_s=37\pm5$ and is within two error bars from
a more recent QMC prediction\cite{rapisard} $r_s=34\pm 4$.  We have
also evaluated the Lindemann ratio $\gamma$, which is shown in
Fig. \ref{linde} as function of $r_s$ near freezing. We find
$\gamma=0.33$ at $r_s=42$, to be compared with an accepted value for
quantum freezing of about $0.3$. We may thus conclude that the SOT is
capable of predicting freezing in two dimensions with good accuracy.

We have also investigated the effect of using a Gaussian ansatz for
the Bloch orbitals (see, e.g., Sec. \ref{gaussian}), as opposed to the
full Kohn-Sham calculation presented above. As expected, this yields
solid energies that are slightly higher, predicting freezing at a
lower density, i.e., at $r_s=50$ with $\gamma=0.25$. Though this is
well out of the range predicted by QMC (see above) it is still a
substantial improvement with respect to an earlier
prediction\cite{deber} of $r_s\agt 80$, using the Gaussian ansatz and
an approximate local field factor.

We should also mention results recently obtained using another DFT
scheme,\cite{ghosh} in which a local density approximation (LDA) to
the total energy of the modulated phase is augmented by gradient
corrections\cite{norman} (GCDFT). This approach yields crystallization
at $r_s=31$, but with a density which appears to be very little
modulated. In fact we have repeated such calculations, solving the
equivalent one-orbital self-consistent Kohn-Sham equations in full. We
reproduce the $r_s=31$ found in Ref. \onlinecite{ghosh} and we find
$\gamma=0.369$ which is very close to the uniform limit of
$\gamma=0.373$. To give a more direct idea of what this means in terms
of localization, we may look at the minimum in the density profile at
half distance between a site and one of its nearest neighbors, in
units of the on-site density, $\delta=n(r_{nn}/2)/n(0)$.  Here
$r_{nn}$ is the nearest neighbor distance.  We find that the GCDFT
predicts $\delta=0.92$ at freezing, to compare with $\delta=0.30$,
which we have obtained within the SOT, and an exact value which is
likely to be even smaller.  With respect to a conventional
LDA\cite{sham,saverio} in which the noninteracting kinetic energy is
treated without approximation, the GCDFT is introducing an
overestimate of the kinetic cost of a modulation. For the small
modulations predicted by the GCDFT, this may easily be checked by
comparing, for instance at the first RLV of the triangular crystal,
the exact $2d$ noninteracting response function\cite{stern,rochester}
with the one corresponding to the GCDFT, $\chi_0^{GC}(q)=-(\sigma
m/2\pi\hbar^2)/[1+q^2/2q_F^2]$.  One might be tempted to argue that,
for small modulations, the approach of Ref. \onlinecite{ghosh} would
be more consistent than a conventional LDA\cite{ldacalc} in that it
treats all the components of the energy on the same footing.  However,
the quality of the resulting density profile, which is indeed very
poor, pointing to a weakly first-order if not a second-order
transition, contradicts such a conclusion.

Returning to the effects of the large-$q$ behavior of the quantum dcf
on freezing we may conclude that these are far less drastic in two
dimensions than in three. Notice, however, that if one tried to apply
the quadratic theory for systems with $r_s > 59$, one would obtain,
also in two dimensions, the erroneous answer that the stable phase is
a crystal with infinite localization. This demonstrates that the
quadratic theory gives a reasonable description of solids whose
thermodynamic parameters are not far away from the freezing ones.
Deeply inside the region of thermodynamic stability of the solid, the
SOT loses its validity, even in those cases where it succeeds in
predicting freezing.

Having concluded the discussion of the quadratic theory in two and
three dimensions, we now proceed with the nonperturbative approach,
i.e. the MWDA.
   
\section{Modified Weighted Density Approximation}
\label{mwdalarge}

In this Section we will examine the behavior and performance of the
MWDA for the case of systems of charged particles. The general
analysis will show that regardless of statistics, the MWDA yields a
functional which is {\it{bounded}} from below, i.e. at the limit of
large localizations the energy difference between the solid and the
liquid tends to $+\infty$ and not $-\infty$ as in the case of the
SOT. Then, we will present the application of the MWDA to the case of
charged bosons, for which the availability of liquid-state input
allows us to perform the MWDA-calculation. We still find the stability
of the crystal to be overestimated, nevertheless.

The presence of a singular term in the dcf $K(r;n_l)$ implies the
existence of a similar term in the weight function $w(r;\hat{n})$, as
is clear from Eqs. (\ref{weight}) and (\ref{regular_r}). With a
straightforward analysis which closely parallels the one developed in
the previous section for the excess energy one is led to the
conclusion that for strong localizations (large values of the
parameter $\alpha$), the weighted density is given to leading order in
$\alpha$ by
\begin{eqnarray}
\hat{n}= \frac{U_0(\hat{n})}{-2\hat{n}\epsilon'(\hat{n})\pi^{d/2}}
\alpha^{d/2},
\label{nsing}
\end{eqnarray}
suggesting that $\hat{n}$ grows with $\alpha$ as we shall demonstrate
shortly, provided that $-\hat{n}\epsilon'(\hat{n})>0$. We have verified
that indeed $-\hat{n}\epsilon'(\hat{n})=
(1/d)\hat{r_s}\epsilon'(\hat{r}_s)>0$ for all the systems considered
below.\cite{tanatar,vosko,conti} We shall examine the behavior of
$\hat{n}$ in two and three space dimensions separately.

\subsection{Charged fermions in two dimensions}

On account of Eq. (\ref{nsing}) above, let us assume that $\hat{n}$
diverges with $\alpha$ and therefore that $\hat{r_s}$---the Wigner
radius in units of Bohr radii corresponding to the effective density
$\hat{n}$---goes to zero in the same limit.  Using the definitions of
the previous section we may eliminate $\hat{n}$ in favor of
$\hat{r}_s$ to obtain
\begin{eqnarray}
\frac{1}{\hat{r}_s^2}= \frac{-(\hat{r}_s\epsilon_c(\hat{r}_s))'}
{\hat{r}_s\epsilon'(\hat{r}_s)}\frac{C}{ r_s^2}\tilde{\alpha},
\label{rs2d}
\end{eqnarray}
with the prime denoting differentiation with respect to $\hat{r}_s$
and $C$ a constant which depends on the structure chosen for the
solid. Here $\epsilon_c$ and $\epsilon$ are respectively the
correlation and the excess energy (exchange plus correlation) of the 2
dimensional electron gas.

For small $r_s$ the excess energy is given by\cite{rapisard}
\begin{eqnarray}
 \epsilon(\hat r_s) = -A\hat r_s^{-1} -B -D \hat r_s \ln \hat r_s +
 ...,
\label{gsen} 
\end{eqnarray}
with $A,B,D$ positive constants which depend on the spin polarization.
The dominant, $-\hat r_s^{-1}$-term is the exchange energy and the
remainder is the correlation energy $\epsilon_c(\hat r_s)$.  Using the
above equation and keeping only the leading terms in the ratio
appearing in Eq. (\ref{rs2d}) one obtains at once
\begin{eqnarray}
\hat{r}_s = (Ar_s^2/BC)^{1/3} \tilde{\alpha}^{-1/3}.
\label{final2d}
\end{eqnarray}
Now, using Eq. (\ref{mw1}) and since the excess energy of the liquid
scales like $-\hat r_s^{-1}$, Eq. (\ref{final2d}) yields
\begin{eqnarray}
   {{E_{ex}^{MWDA}(r_s,\tilde\alpha)}\over{N}}\Bigg|_{\tilde\alpha \gg
   1} = - |E_{2d}(r_s)| \tilde\alpha^{1/3}. \label{scalen2d}
\end{eqnarray}
Since the ideal term scales as $\tilde\alpha$, it dominates over the
excess one at large $\tilde\alpha$ and thus the total energy tends to
plus infinity at the limit of strong localization. Notice that in the
SOT the excess term scales as $-\tilde\alpha$, whereas here only as
$-\tilde\alpha^{1/3}$. The MWDA makes the dependence of the excess
energy on the localization parameter a lot weaker than the SOT. We
will see that this is also the case in three dimensions.

\subsection{Charged bosons and fermions in three dimensions}

In three dimensions, eliminating $\hat{n}$ in favor of $\hat{r}_s$ in
Eq. (\ref{nsing}) yields
\begin{eqnarray}
\frac{1}{\hat{r}_s^3}= \frac{-(\hat{r}_s\epsilon_c(\hat{r}_s))'}
{\hat{r}_s\epsilon'(\hat{r}_s)}\frac{C}{ r_s^3}\tilde{\alpha}^{3/2},
\label{rs3d}
\end{eqnarray}
where $\epsilon_c$ and $\epsilon$ have the usual meaning for fermions,
whereas for bosons $\epsilon_c$ coincides with $\epsilon$---the total
energy per particle. Again, $C$ is a constant depending on the
structure assumed for the solid. As in two dimensions, we are led to
assume that $\hat{r}_s\to 0$ as $\tilde{\alpha}\to\infty$ and
therefore we shall retain in this limit only leading terms in
Eq. (\ref{rs3d}).

{\it{Charged bosons}}. As we have already mentioned
$\epsilon_c(\hat{r}_s)=\epsilon(\hat{r}_s)$ in this case and as $\hat
r_s \to 0$, we have to dominant order $\epsilon(\hat r_s) = -0.8031
\hat r_s^{-3/4}$ Ry.\cite{conti,brueckner} Thus we obtain
\begin{eqnarray}
\hat{r}_s=(3/C)^{1/3}r_s\tilde{\alpha}^{-1/2}. 
\end{eqnarray}
The energy $\epsilon(\hat{r}_s)$ now scales as $-\hat r_s^{-3/4}$,
thus
\begin{eqnarray}
   {{E_{ex}^{MWDA}(r_s,\tilde\alpha)}\over{N}}\Bigg|_{\tilde\alpha \gg
   1} = -|E_b(r_s)| \tilde\alpha^{3/8}. \label{scalenbos}
\end{eqnarray}
The MWDA-excess energy per particle scales only as
$-\tilde\alpha^{3/8}$ as opposed to $-\tilde\alpha^{3/2}$ in the
SOT. Thus, the ideal energy which is linear in $\tilde\alpha$
dominates for strong localizations, and the MWDA-functional is free of
the pathology of the SOT, i.e. it does have a lower bound.

The actual calculations that we carried out were for the case of
bosons only; however, the same analysis can be carried out for
fermions, and the results for this case are presented below.

{\it{Charged fermions}}. The first few terms in the expansion of the
excess ground-state energy of the electron fluid for small $\hat r_s$
read as \cite{vosko}
\begin{eqnarray}
   \tilde \epsilon(\hat r_s) = 
   -B \hat r_s^{-1} + \Gamma \ln \hat r_s - \Delta + 
   O(\hat r_s), \label{hf}
\end{eqnarray}
where all constants are positive. Once more, the term proportional to
$-\hat r_s^{-1}$ is the exchange energy and the remainder is the
correlation energy.  Using Eqs. (\ref{rs3d}) and (\ref{hf}) we obtain
as $\hat{r}_s\to 0$
\begin{eqnarray}
\frac{1}{\hat{r}_s^3}= \hat{r}_s|\ln{\hat{r}_s}|\frac{\Gamma C}{B}
\frac{1}{r_s^3} \tilde{\alpha}^{3/2}.
\label{rsfer1}
\end{eqnarray}
To leading order as $\tilde\alpha\to\infty$ we immediately obtain
\begin{eqnarray}
\hat{r}_s = D(r_s)\frac{\tilde\alpha^{-3/8}}{[\ln\tilde\alpha
]^{1/4}},
\label{rsfer}
\end{eqnarray}
with $D(r_s)= [8Br_s^3/3\Gamma C]^{1/4}$.  Finally, since the excess
energy per particle scales as $-\hat r_s^{-1}$, the MWDA-functional
obeys the scaling
\begin{eqnarray}
   {{E_{ex}^{MWDA}(r_s,\tilde\alpha)}\over{N}}\Bigg|_{\tilde\alpha \gg
   1} = -|E_f(r_s)| [\ln \tilde\alpha]^{1/4} \tilde\alpha^{3/8}.
\label{scalenfer}
\end{eqnarray}
Thus, the MWDA is free of the unboundedness problem also for fermions.
It is interesting that the scaling of the excess energy is now
dependent of the statistics [see Eqs. (\ref{scalenbos}) and
(\ref{scalenfer}) above], though very weakly, due to the logarithmic
dependence in $\tilde\alpha$ present for fermions. This is at variance
with the prediction of the SOT, where the same scaling was found and
it appears intriguing. In fact, naive considerations would suggest
that the excess energy should scale in the same way for bosons and
fermions at the strong localization limit since, in this case, each
particle is confined to its own cell and statistics becomes
unimportant.

The existence of a lower bound for the MWDA-functional is an
improvement over the behavior of the SOT. However, this property
guarantees neither the existence of a minimum at nonzero localization
nor its correct location and behavior in terms of changes of the
average density. If, for example, the total energy is monotonically
increasing as a function of $\alpha$, then the only minimum will occur
for the uniform liquid. On the other hand, it is possible that a
minimum always exists, for any value of the average density and is
lower than the liquid one; in this second case, we are led to the
erroneous prediction that the crystal is stable at all densities. In
the following subsection we show the results of the full
MWDA-calculation for charged bosons and we find that, in fact, this
second scenario materializes.

\begin{figure}
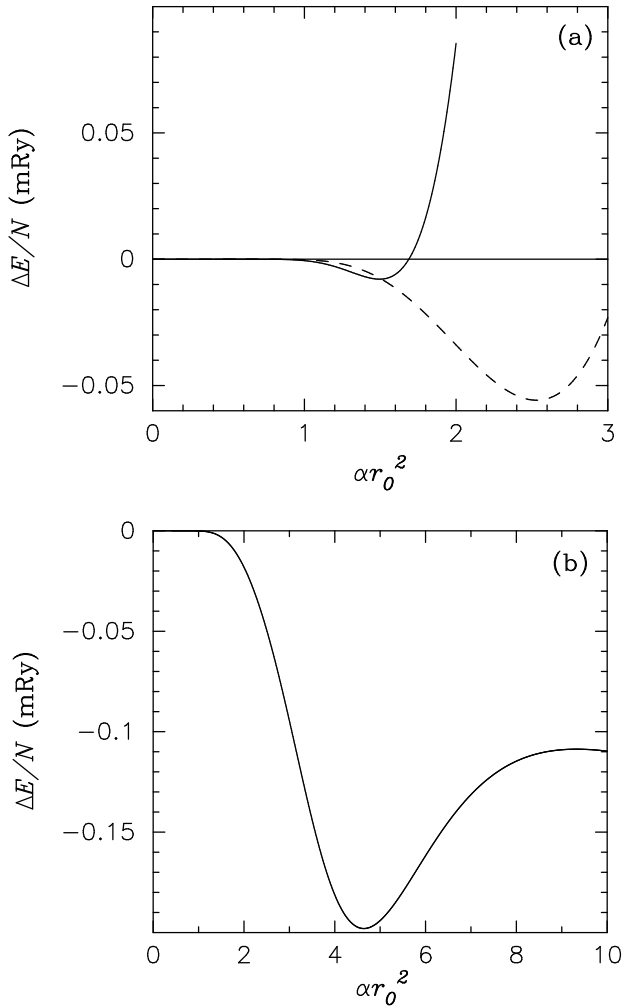

\null
\vspace{-3.1cm}
\hspace{.3cm}\psfig{figure=denmw.ps,width=6.5cm}
\vbox{\vspace{-1.2cm}\hspace{0.2cm}\psfig{figure=denmw2.ps,width=6.5cm}}
\vspace{1.1cm}
\caption[dum05] {Ground-state energy difference\cite{explain}
between the the charged boson
bcc-solid and the liquid, 
versus localization at three different average densities,
as obtained in the MWDA. (a) Solid line: $r_s = 20$; dashed line: $r_s
= 50$. (b) $r_s = 100$.}
\label{denmw}
\end{figure}

\subsection{MWDA-calculation for charged bosons}

We have implemented the MWDA-self-consistency condition
[Eq. (\ref{mw2})] for the case of charged bosons for which there exist
sufficient simulation data for the local field factor for a range of
densities varying from $r_s = 10$ to $r_s = 160$
(Ref. \onlinecite{conti}).  We have used an analytic fit to the
equation of state obtained from simulation.\cite{conti} We limit our
study to the charged boson liquid because for polarized fermions the
only available Monte Carlo data are for $r_s = 100$ and the
implementation of the MWDA requires the knowledge of the dcf of the
liquid over a wide range of densities.

We have carried out the MWDA-calculation with the Gaussian ansatz for
three different values for the average density, namely $r_s = 20$,
$50$ and $100$. The results are shown in Fig.~\ref{denmw}.  It can be
seen immediately that, unlike the SOT, the MWDA gives minima of
$\Delta E/N$ for finite values of the localization parameter
$\alpha$. Moreover, the trends of these minima are correct: they get
deeper and also move towards stronger localization as the average
density decreases. However, according to simulations \cite{alder} the
liquid is stable for $r_s < 160 \pm 10$, whereas the MWDA gives a
{\it{lower}} energy for the bcc-solid for values of $r_s$ as low as
20.  Thus we can say that although the MWDA is already much better
than the SOT, it still predicts a solid that is {\it{too
stable}}. Even at high densities the MWDA-functional has a global
minimum for a modulated phase.

\section{Conclusions}

The implementation of the correct liquid-state input in a
density-functional approach to the freezing of quantum liquids brings
about a remarkable new result, namely that {\it{in three dimensions}}
the standard SOT suffers from a lack of a lower bound and is always
{\it strictly} minimized for infinitely localized solids. As we have
already mentioned, one might argue that the perturbative character of
the SOT limits its validity to weakly modulated density profiles and
thus local minima for finite localization should suffice. However, as
we have demonstrated above for $^4$He and charged bosons, such
minima---when they exist---still yield an incorrect description of
freezing, predicting stability of the solid well inside the region
where the system in fact is liquid. Recourse to non-perturbative
theories including certain classes of higher order terms, such as the
MDWA, does not help much in practice. One gets rid of the extreme
pathology of the SOT theory in that the resulting functional is
bounded from below, which is certainly satisfactory. However even the
MDWA predicts the crystal to be the stable phase deep into the region
where the liquid should be stable.

The asymptotic analysis in the localization parameter $\tilde\alpha$
that we have carried out above for the MDWA is easily generalized to
the weighted density approximation (WDA),\cite{curtin,saverio} once it
is realized that the weight function $\tilde{w}(q; n)$ has in this
case the same large $q$ limit as in MWDA, and therefore the same
singular term in real space. One obtains the same scalings as
discussed in Sec. \ref{mwdalarge} above, and therefore bounded
functionals. Whether the predictions of the WDA for freezing will be
any better than those of the MWDA remains to be investigated, though
we doubt it. We may mention that the conventional
LDA\cite{sham,saverio} also brings about bounded functionals, as one
obtains almost at once. The scaling is the same as for the MWDA for electrons
in 2 dimensions and charged bosons in 3 dimensions; it is different
for electrons in 3 dimensions, for which the exchange-correlation
energy goes like $-\tilde\alpha^{1/2}$.  Again, the LDA is making the
solid too stable in three dimensions.\cite{senatore}
 
The situation in two dimensions appears specular to the one summarized
above for three dimensions. In fact, earlier applications of the DFT
theory of freezing with approximate liquid dcf gave good results in
three dimensions,\cite{senatore,denton} while failing in two
dimensions.\cite{deber} We have demonstrated above that the reverse is
true, if more accurate liquid input is used, which obeys the exact
large $q$ behavior discussed in Sec. \ref{dcfq}. In particular, in two
dimensions the SOT provides bounded functionals, in the relevant
region of density, and yields a good description on the freezing
transition.

The recently obtained accurate information on the liquid-state linear
response functions gives rise, therefore, to a new problem: our
favorite approximate density-functional schemes which we have learned
to trust from our experience on classical systems, seem to fail when
applied to quantum systems in three dimensions. There is a need for
reexamination of the current formalism of quantum density-functional
theory of freezing and the development of approximate schemes which
will be more appropriate to deal with the peculiarities of quantum
systems. In this respect, the use of direct correlation functions
possessing the correct asymptotic behavior is crucial, as is such a
behavior that causes all the present troubles. At variance with the
classical case we have seen that quantum functionals tend to predict
excess energies (per particle) that negatively diverge at infinite
localization. Though the MWDA produces a functional bounded from
below, we speculate that the divergence of its excess part in this
limit could still be incorrect, as the potential energy should remain
finite, unless one can prove that it is the kinetic contribution to
the excess energy which is bringing about this divergence.

\section*{ACKNOWLEDGMENTS}
We would like to thank Professor S. Stringari both for helpful
discussions and for sending us unpublished material and
Dr. A. R. Denton for sending us a copy of Ref. \onlinecite{preprint}
prior to publication. Two of us (S.M. and G.S.) also acknowledge fruitful
collaboration with Dr. A. Debernardi in early work on the freezing of
the $2d$ electron gas.  C.N.L. has been supported by the Human Capital
and Mobility Programme of the Commission of the European Communities,
Contract No.  ERBCHBICT940940.


\begin{references}
\bibitem{ldacalc}We have performed a conventional LDA calculation
treating the noninteracting kinetic energy without approximations and
we find freezing from the polarized $2d$ electron gas into the
triangular lattice at an $r_s$ as low as 15, with $\delta=0.40$.
\end{references}
\end{document}